\documentclass[12pt,letterpaper]{IEEEtran}
\onecolumn

\usepackage{cite}
\usepackage{amsthm}
\usepackage{amssymb,amsmath,bm}
\usepackage{multirow}
\usepackage{macros}
\usepackage{times}
\usepackage{graphicx}
\usepackage{epsfig}
\usepackage{color}
\usepackage{array}
\usepackage{tabularx}
\usepackage{algorithm}
\usepackage{algorithmic}

\renewcommand{\baselinestretch}{2.0}

\begin{document}

\title{Constrained Function Based En-Route Filtering for Sensor Networks}

\author{Chia-Mu~Yu,~\IEEEmembership{Student Member,~IEEE,}
        Chun-Shien~Lu,~\IEEEmembership{Member,~IEEE,}
        and~Sy-Yen~Kuo,~\IEEEmembership{Fellow,~IEEE}% <-this % stops a space
\IEEEcompsocitemizethanks{
\IEEEcompsocthanksitem Chia-Mu Yu (r91045@csie.ntu.edu.tw) is with National Taiwan University and Academia Sinica. Chun-Shien Lu (lcs@iis.sinica.edu.tw) is with Academia Sinica. Sy-Yen Kuo (sykuo@cc.ee.ntu.edu.tw) is with National Taiwan University. Corresponding author: Chun-Shien Lu}.
}

\date{}
\maketitle

%%%%%%%%%%%%%%%%%%%%%%%%%%%%%%%%%%%%%%%%%%%%%%%%%%%%%%%%%%%%%
\begin{abstract}
Sensor networks are vulnerable to \emph{false data injection attack} and \emph{path-based DoS} (PDoS) attack. While conventional authentication schemes are insufficient for solving these security conflicts, an \emph{en-route filtering} scheme acts as a defense against these two attacks. To construct an efficient en-route filtering scheme, this paper first presents a Constrained Function based message Authentication (CFA) scheme, which can be thought of as a hash function directly supporting the en-route filtering functionality. Together with the \emph{redundancy property} of sensor networks, which means that an event can be simultaneously observed by multiple sensor nodes, the devised CFA scheme is used to construct a CFA-based en-route filtering (CFAEF) scheme. In contrast to most of the existing methods, which rely on complicated security associations among sensor nodes, our design, which directly exploits an en-route filtering hash function, appears to be novel. We examine the CFA and CFAEF schemes from both the theoretical and numerical aspects to demonstrate their efficiency and effectiveness.
\end{abstract}

{\indent\bf keywords\/}: Sensor Networks, Authentication, En-route Filtering, Security
\clearpage
%%%%%%%%%%%%%%%%%%%%%%%%%%%%%%%%%%%%%%%%%%%%%%%%%%%%%%%%%%%%%
\section{Introduction}\label{sec: Introduction}
A Wireless Sensor Network (WSN) is composed of a large number of sensor nodes with limited resources. Since WSNs can be deployed in an unattended or hostile environment, the design of an efficient authentication scheme is of great importance to the data authenticity and integrity in WSNs. In this respect, many authentication schemes have been proposed. The most straightforward way to guarantee data authenticity is to use conventional public-key cryptography based digital signature techniques. Although the use of public-key cryptography on WSNs has been demonstrated in \cite{ln08,mws08} to be feasible, the computation overhead is still rather high for resource-constrained devices.

\textbf{Authentication Problem.} Sensor networks are vulnerable to \emph{false data injection attack} \cite{ylk09}, by which the adversary injects false data, attempting to either deceive the base station (BS, or data sink), and \emph{path-based DoS} (PDoS) attack \cite{dhm05}, by which the adversary sends bogus messages to randomly selected nodes so as to waste the energy of forwarding nodes\footnote{The terms ``forwarding node'' and ``intermediate node'' are used interchangeably in this paper.}. Several so-called \emph{en-route filtering schemes} have been proposed to quickly discover and remove the bogus event report injected by the adversary. Here, ``en-route filtering'' means that not only the destination node but also the intermediate nodes can check the authenticity of the message in order to reduce the number of hops the bogus message travels and, thereby, conserve energy. Hence, it is especially useful in mitigating false data injection attack and PDoS attack \cite{dhm05}, because the falsified messages will be filtered out as soon as possible.

\textbf{Related Work.} SEF \cite{yllz04} is the first en-route filtering scheme found in the literature that exploits probabilistic key sharing over a partitioned key pool. Due to its design strategy, however, only a few intermediate nodes between the source-destination node pair have the ability to check the validity of forwarding messages, leading to low filtering capability. IHA \cite{zsjn04}, which verifies the transmitted packets in a deterministic hop-by-hop fashion, has also been proposed to authenticate the event report. Nevertheless, it requires complicated key sharing among neighboring nodes and could be vulnerable to node compromises if node compromises are mounted immediately after sensor deployment. Based on the similar idea used in SEF and IHA, several other en-route filtering schemes are proposed. With the sophisticated use of one-way hash chains in clustered sensor networks, DEF \cite{yg06} has improved filtering power over SEF \cite{yllz04}. Using the proposed multiple-axis technique, GREF \cite{yl09} is designed to support en-route filtering in the networks with multiple data sinks. LBRS \cite{yyyla05} and LEDS \cite{rlz06} take advantage of location information to enhance the resilience to node compromises. CCEF \cite{yl04}, STEF \cite{ksbe07}, and KAEF \cite{yzzm08} are presented to authenticate the transmitted packets only in query-based sensor networks.

Note that, as to broadcast authentication, $\mu$TESLA and its variants \cite{ln04,pswct01} can also serve message authentication well. Nevertheless, broadcast authentication is used to authenticate only the messages sent from the base station while en-route filtering schemes are used for authenticating and filtering a bogus event report that is assumed to not be detected by multiple legitimate sensor nodes in a node-to-node or node-to-BS communication pattern. Thus, the design of broadcast authentication schemes is orthogonal to the content of this paper.

\textbf{The Design of En-Route Filtering Schemes.} The redundancy property, which means that an event can be simultaneously observed by multiple sensor nodes, can be used to design the en-route filtering schemes. Specifically, the general design framework is that the source node that senses an event and wants to send an event report to the destination node first collects the neighboring nodes' endorsements of the sensed event. Afterwards, it sends out the event report and endorsements. Each intermediate node and the destination node can check the authenticity of the received report via the verification of the endorsements.

Aiming to enhance the filtering capability and improve the resilience against node compromises, most of the existing en-route filtering schemes rely on complicated security associations (\emph{e.g.} key sharing), and, therefore, incur some assumptions such as secure bootstrapping time, stable routing, single data sink, the immobility of sensor nodes, \emph{etc}, making them impractical. We identify the following four problems associated with the existing schemes.
\begin{enumerate}
\item The reason the unnecessary assumptions should be made stems from the fact that the message authentication codes (MACs, or keyed hash functions) used do not support en-route filtering functionality, while the authenticity of the forwarding messages needs to be checked by as many intermediate nodes as possible.
\item It has been demonstrated in \cite{cskm05,dghm05,sy07} that the node is able to send an event report to the other nodes in certain in-network control scenarios. Nonetheless, the existing schemes, which are only effective on the node-to-BS communication pattern, are ineffective in handling false data injection and PDoS attacks in such scenarios.
\item The existing en-route filtering schemes are difficult to apply on mobile sensor networks or networks with multiple sinks. In other words, the applicability of en-route filtering schemes on different network settings should be improved.
\item Last, based on conventional design, all the en-route filtering schemes suffer from a special kind of DoS attack, false endorsement DoS (FEDoS) attack \cite{kse08}, which could neutralize the advantages gained from the use of en-route filtering schemes.
\end{enumerate}

In this paper, we take a completely different approach to the design of an en-route filtering scheme to avoid the above problems. In particular, instead of establishing security associations, we turn to construct an en-route filtering hash function, Constrained Function-based Authentication (CFA) scheme, and then employ such hash function to generate MACs used to endorse the sensor readings so that each intermediate node can verify the authenticity of forwarding messages. In particular, our proposed CFA possesses the following four characteristics: 1) \emph{Resilience to node compromise} (RNC), which means that the compromised nodes cannot forge the messages sent from the genuine nodes; 2) \emph{Immediate authentication} (IA), which can be thought of as a synonym to en-route filtering and can be used to filter out the falsified messages as soon as possible to conserve energy; 3) \emph{Independence to network setting} (INS), which means that CFA can be applied on the networks with different network settings; 4) \emph{Efficiency} (EFF), which means that CFA has low computational and communication overhead. With these characteristics, a CFA-based en-route filtering (CFAEF) scheme can be constructed in such a way that the source node sends to the destination node a message, together with the corresponding CFA-based endorsements generated by the neighboring nodes. Afterwards, the source node can determine if the neighboring nodes send the false endorsement and each intermediate node has ability to check the authenticity of forwarding messages. As a whole, as we will show later, the advantages of applying CFA on MAC generation are that the filtering capability can be improved, the resilience against FEDoS attack can be achieved, and the impractical assumptions previously made in the literature are no longer required.

\textbf{Our Contributions.} Our contributions are as follows:
\begin{itemize}
\item A Constrained Function based Authentication (CFA) scheme for WSNs is proposed. CFA can be thought of as a hash function directly supporting en-route filtering functionality, and can act as a building block for other security mechanisms.

\item A CFA-based En-route Filtering (CFAEF) scheme that can simultaneously defend against false data injection, PDoS, and FEDoS attacks is proposed. Particularly, compared with the existing methods, which either have low filtering capability or necessitate some unrealistic assumptions, our CFAEF scheme can be applied on arbitrary networks without further assumptions.

\item The efficiency of CFA and CFAEF schemes is studied in both theoretical and numerical aspects.
\end{itemize}

\section{System Model}
\textbf{Network model.} We assume a WSN composed of $N$ resource-limited sensor nodes with IDs, $\mathcal{I}\subset \mathbb{N}$. The unique ID for each node can be either arbitrarily assigned in the sensor platform, such as telosB, or fixed in a specific sensing hardware when manufactured, like the MAC address on current Network Interface Cards (NICs). Although one or multiple base stations (or data sinks) are involved in data collection in a WSN, the efficiency of our proposed schemes does not rely on their trustworthiness and authenticity. In addition, arbitrary network topology is allowed in our method. Some or all of the sensor nodes can have mobility. The network planner, prior to sensor deployment, also cannot gain any deployment knowledge pertaining to sensors' locations.

\textbf{Security model.} The objectives of the adversary are to deceive the BS into accepting the falsified event report and to deplete sensor nodes' energy by launching PDoS attack and FEDoS attack. In this paper, sensor nodes are assumed to not be equipped with tamper-resistant hardware. Thus, all the information is exposed and can be utilized by the adversary as long as a node is captured. We also assume that the attacks such as node compromises can be mounted by the adversary immediately after sensor deployment, \emph{i.e.}, the proposed schemes cannot rely on the secure bootstrapping time used in \cite{rlz06,zsj03}. If required, any pair of sensor nodes can establish their shared key\footnote{Here, the key establishment scheme in \cite{ylk09}, instead of the ones in \cite{cps03,cy07,ddhv06,eg02,lcmx08,lnl05}, is chosen to be used in our proposed method because the latter are \emph{interactive}, which means that two nodes require to communicate with each other once they would like to establish their common key.} in a noninteractive fashion \cite{ylk09}. Although sensor networks are known to be vulnerable to many attacks such as wormhole attack, selective forwarding attack, \emph{etc}, we refer to the existing rich literature \cite{cch07,hpj06,kw03,wwxwa08} for these issues and the defense against these attacks is beyond the scope of this paper.

\section{The Constrained Function Based Authentication (CFA) Scheme}\label{sec: The Constrained Function Based Authentication (CFA) Scheme}
Since the proposed CFA scheme is constructed by making use of the pairwise key generated by the CARPY+ scheme \cite{ylk09} for secure communication, we first briefly review CARPY+ in Sec. \ref{sec: A Review of CARPY+ Scheme} to make this paper self-contained. Then, the proposed CFA scheme will be presented in the remaining subsections. In this paper, nodes $u$, $v$, and $\varepsilon$ are denoted as the source node, destination node, and intermediate node, respectively.

\subsection{Review of the CARPY+ Scheme \cite{ylk09}}\label{sec: A Review of CARPY+ Scheme}
Let $N$, $\lambda$, and $\mathbb{F}_q=\{0,\dots,q-1\}$, where $q$ is a prime number, be the number of sensor nodes, a security parameter independent of $N$, and a finite field, respectively. Let $A=(D\cdot G)^T$, where $D\in \mathbb{F}_q^{(\lambda+1)\times (\lambda+1)}$ is a symmetric matrix, $G\in \mathbb{F}_q^{(\lambda+1)\times N}$ is a matrix, and $(D\cdot G)^T$ is the transpose of $(D\cdot G)$. Let $K=A\cdot G$. It can be known that $K$ must be symmetric because $A\cdot G=(D\cdot G)^T\cdot G=G^T\cdot D\cdot G=(A\cdot G)^T$. Before sensor deployment, proper constrained random perturbation vectors are selected and applied on each row vector of $A$ to construct a matrix $W$. In addition, $G$ is selected as a Vandermonde matrix generated by a seed. The $j$-th row vector of $W$, $W_{j,-}$, is stored into the node $j$. After sensor deployment, node $u$ can have the shared key with node $v$ by calculating the inner product of the row vector $W_{u,-}$ and the $v$-th column vector $G_{-,v}$, then extracting the common part as the shared key. Note that in the CARPY+ scheme, $G$ and $s$ can be publicly known while $A$ should be kept secret. Therefore, CARPY+ can establish a pairwise key between each pair of sensor nodes without needing any communication. This property is an essential part in constructing the proposed CFA scheme, because establishing a key via communications incurs the authentication problem, leading to a circular dependency.

\subsection{Basic Idea}\label{sec: Basic Idea}
In the CFA scheme, the network planner, before sensor deployment, selects a secret polynomial $f(x,y,z,w)$ from the set $\mathfrak{F}$ (to be defined in Eq. (\ref{eq: Fdefinition}) later), whose coefficients should be kept as secret, thereby constituting the security basis of CFA. For simplicity, we assume that the degree of each variable in $f(x,y,z,w)$ is the same, which is $d$, although they can be distinct in our scheme. For each node $u$, the network planner constructs two polynomials, $f_{u,1}(y,z,w)=f(u,y,z,w)$ and $f_{u,2}(x,z,w)=f(x,u,z,w)$. Since directly storing these two polynomials enables the adversary to obtain the coefficients of $f(x,y,z,w)$ by capturing a few nodes, the authentication polynomial $auth_u(y,z,w)$ and verification polynomial $verf_u(x,z,w)$ should be, respectively, constructed from the polynomials $f_{u,1}(y,z,w)$ and $f_{u,2}(x,z,w)$ by adding independent perturbation polynomials. Afterwards, the authentication and verification polynomials, instead of $f_{u,1}(y,z,w)$ and $f_{u,2}(x,z,w)$, are stored in node $u$. For source node $u$, the MAC attached to the message $m$ is calculated according to its own authentication polynomial. Let \emph{verification number} be the result calculated from the verification polynomial $verf_u(x,z,w)$ by substituting the claimed source node ID, the shared pairwise key, and the hashed message into $x$, $z$, and $w$, respectively. The received node considers the received message authentic and intact if and only if the \emph{verification difference}, which is the difference between the received MAC and its calculated verification number, is within a certain predetermined range.

Although our CFA scheme is similar to Zhang \emph{et al}.'s scheme \cite{zsw08}, the design strategies used in the CFA scheme are different from the ones in \cite{zsw08}, except the fact that both rely on polynomial evaluation. In Zhang \emph{et al}.'s scheme, due to the improper use of perturbation, the nodes' IDs should be forced to be changed, resulting in the limitation of hardware dependence. In addition, as an arbitrary secret polynomial can be used in \cite{zsw08}, immediate authentication can be achieved only if the message authentication code forms a polynomial. On the contrary, since the secret polynomial $f(x,y,z,w)$ in CFA is selected such that certain properties are satisfied, the message authentication code can be reduced from a polynomial size to a single number, resulting in less communication overhead (packet overhead). On the other hand, whereas the pairwise key has been considered useless in providing either immediate authentication or resilience to node compromises in previous methods, in this paper we find that the pairwise key is helpful in enhancing the security while retaining the property of immediate authentication. Hence, all these characteristics substantially differentiate CFA from \cite{zsw08}.

In the following two subsections, the off-line step and on-line step, respectively, will be described.

\subsection{Off-line Step of CFA scheme}\label{sec: Off-line Step of CFA scheme}
Before deploying sensor nodes, the network planner picks a parameter $q$ from which a finite field $\mathbb{F}_q$ is built. All of the operations throughout the paper are performed over $\mathbb{F}_q$ unless specifically mentioned. Let $\mathcal{I}$ be the set of node IDs. Let $\ell$ be the least number of bits sufficient to represent $q$. Assume that node IDs, pairwise key, and hash value can be represented in $\mathbb{F}_q$. In addition, a security parameter $r<\ell$ is also selected. Then, the secret polynomials $f(x,y,z,w)$'s, used as the basis for constructing both authentication and verification polynomials, are defined in \emph{constrained function set}, $\mathfrak{F}$, where
\begin{align} \mathfrak{F}=\big\{&f(x,y,z,w)| |f(x,y,z,w)-f(x,y',z',w)|\leq 2^{r-1}, |f(x,y,z,w)-f(x',y',z',w)|\geq 3\cdot 2^{r-1}-1,\notag \\
&|f(x,y,z,w)-f(x',y',z',w')|\geq 3\cdot 2^{r-1}-1, x,y\in\mathcal{I}, x'\neq x, y'\neq y, z'\neq z, w'\neq w, r<\ell\big\}.\label{eq: Fdefinition}
\end{align}
The authentication polynomial, $auth_u(y,z,w)=f(u,y,z,w)+n_{u,\mathfrak{a}}(y,z)$, and verification polynomial, $verf_u(x,z,w)=f(x,u,z,w)+n_{u,\mathfrak{v}}(x,z)$, are stored in each node $u$, where polynomials $n_{u,\mathfrak{a}}(y,z)$ and $n_{u,\mathfrak{v}}(x,z)$, used for perturbation, are randomly selected from the \emph{authentication perturbation set}, $\mathfrak{N_a}=\{n(y,z)| 0\leq n(y,z)\leq 2^{r-2}-1, y\in \mathcal{I}, 0\leq y,z\leq q-1\},$ and the \emph{verification perturbation set}, $\mathfrak{N_v}=\{n(x,z)| 0\leq n(y,z)\leq 2^{r-1}-1, x\in \mathcal{I}, 0\leq x,z\leq q-1\},$ respectively. Though the sets $\mathfrak{F}$, $\mathfrak{N_a}$, and $\mathfrak{N_v}$ appear to be artificial, they guarantee the efficiency and feasibility of immediate authentication of CFA. In addition, constructing $auth_u(y,z,w)$ and $verf_u(x,z,w)$ from $\mathfrak{F}$, $\mathfrak{N_a}$, and $\mathfrak{N_v}$ may be time- and energy-consuming. Nevertheless, it could be acceptable because such construction is performed only by the network planner, instead of sensor nodes. If the time required for constructing $auth_u(y,z,w)$ and $verf_u(x,z,w)$ is still an issue that cannot be ignored, an efficient method for constructing the polynomials in a restricted version of $\mathfrak{F}$ will be later discussed in Sec. \ref{sec: Implementation Issues}. The off-line procedure of CFA is described in Fig. \ref{algo: Off-line Step of CFA scheme}.

\begin{figure}[hbt]
\begin{center}
\noindent  \hspace*{-0.1in} \fbox{ \parbox{10cm} {
\begin{myalgorithm}
\footnotesize
123\=123\=456\=789\=012\=345\=678\=901\=234\=567\=890\=123\=456\=789
\kill
{\bf Algorithm:} CFA-Off-line-Step($q$, $r$) \\
1.\enspace Randomly picks a secret polynomial $f(x,y,z,w)\in \mathfrak{F}$\\
2.\enspace \bfor each node $u$\\
3.\enspace\enspace Randomly picks $n_{u,\mathfrak{a}}(y,z)\in \mathfrak{N_a}$ and $n_{u,\mathfrak{v}}(y,z)\in \mathfrak{N_v}$\\
4.\enspace\enspace Store $auth_u(y,z,w):=f(u,y,z,w)+n_{u,\mathfrak{a}}(y,z)$\\
5.\enspace\enspace Store $verf_u(x,z,w):=f(x,u,z,w)+n_{u,\mathfrak{v}}(x,z)$
\end{myalgorithm}
} \noindent } \caption{Off-line Step of CFA.}
\label{algo: Off-line Step of CFA scheme}
\end{center}
\end{figure}

\subsection{On-line Step of CFA scheme}\label{sec: On-line Step of CFA scheme}
After sensor deployment, the sensor node may work as a source node, intermediate node, or destination node depending on whether the message is to be sent or verified. In the following, we describe the operations one should perform when the node acts as different roles. It should be noted that the pairwise key $K_{u,v}=K_{v,u}$, used here, is constructed by applying the CARPY+ scheme \cite{ylk09} on nodes $u$ and $v$, respectively.

\textbf{Source node (Message transmission).} When node $u$ wants to send a message $m$ to node $v$, it calculates the message authentication code: \[MAC_u(v,m)=auth_u(v,K_{u,v},h(m))+n_{u,s},\] where $n_{u,s}$ is randomly picked from the set $\{0,\dots,2^{r-2}\}$. Then, the packet $\mathcal{M}=\langle u,v,m,MAC_u(v,m)\rangle$ is sent to $v$ possibly through a multi-hop path. Note that the message authentication code $MAC_u(v,m)$ is only a number here.

\textbf{Destination node (Message verification).} After receiving the packet $\mathcal{M}=\langle u,v,m,MAC_u(v,m)\rangle$, the destination node $v$ first calculates the verification number: \[verf_v(u,K_{v,u},h(m)),\] according to its own verification polynomial $verf_v(x,z,w)$ and then calculates the corresponding \emph{verification difference}, $VD_{v,u}$: \[VD_{v,u}=|verf_v(u,K_{v,u},h(m))-MAC_u(v,m)|.\] If $VD_{v,u}$ is within the range $\{0,\dots,2^{r-1}-1\}$, where $r$ is a security parameter mentioned in Sec. \ref{sec: Off-line Step of CFA scheme}, then the authenticity and integrity of the packet $\mathcal{M}$ is successfully verified. Otherwise, the packet $\mathcal{M}$ is dropped. The principle behind this step is as follows: \begin{align} &verf_v(u,K_{v,u},h(m))-MAC_u(v,m)\notag\\
=&(f(u,v,K_{v,u},h(m))+n_{v,\mathfrak{v}}(u,K_{v,u}))-(f(u,v,K_{u,v},h(m))+n_{u,\mathfrak{a}}(v,K_{u,v})+n_{u,s})\notag\\
    =&(f(u,v,K_{v,u},h(m))-f(u,v,K_{u,v},h(m)))+(n_{i,\mathfrak{v}}(u,K_{v,u})-(n_{u,\mathfrak{a}}(v,K_{u,v})+n_{u,s})\notag\\
    =&n_{i,\mathfrak{v}}(u,K_{v,u})-(n_{u,\mathfrak{a}}(v,K_{u,v})+n_{u,s}).\label{eq: destination_verf}\end{align}
>From the rules of constructing authentication and verification polynomials, we know that $n_{i,\mathfrak{v}}(u,K_{i,u})\in \{0,\dots,2^{r-1}-1\}$, $n_{u,\mathfrak{a}}(v,K_{u,v})\in \{0,\dots,$ $2^{r-2}-1\}$, and $n_{u,s}\in \{0,\dots,2^{r-2}\}$. Thus, when $\mathcal{M}$ is genuine, the verification difference $VD_{v,u}=|verf_v(u,K_{v,u},h(m))-MAC_u(v,m)|$ must be within $\{0,\dots,2^{r-1}-1\}$.

\textbf{Intermediate node (Message verification).} After receiving the packet $\mathcal{M}=\langle u,v,m,MAC_u(v,m)\rangle$, the intermediate node $\varepsilon$ first calculates $verf_\varepsilon(u,K_{i,u},h(m))$ according to its own verification polynomial $verf_\varepsilon(x,z,w)$ and then calculates the verification difference $VD_{\varepsilon,u}=|verf_\varepsilon(u,K_{\varepsilon,u},h(m))-MAC_u(v,m)|$. If $VD_{\varepsilon,u}$ is within the range $\{0,\dots,2^r-1\}$, then the authenticity of the packet $\mathcal{M}$ is successfully verified, and the packet $\mathcal{M}$ will be forwarded by node $\varepsilon$. Otherwise, the packet $\mathcal{M}$ is dropped. The principle behind this step is as follows. When a genuine packet $\mathcal{M}$ is received, we can obtain: \begin{align} &verf_\varepsilon(u,K_{\varepsilon,u},h(m))-MAC_u(v,m)\notag\\
 =&(f(u,\varepsilon,K_{\varepsilon,u},h(m))+n_{\varepsilon,\mathfrak{v}}(u,K_{\varepsilon,u}))-(f(u,v,K_{u,v},h(m))+n_{u,\mathfrak{a}}(v,K_{u,v})+n_{u,s})\notag\\
    =&(f(u,\varepsilon,K_{\varepsilon,u},h(m))-f(u,v,K_{u,v},h(m)))+(n_{\varepsilon,\mathfrak{v}}(u,K_{i,u})-(n_{u,\mathfrak{a}}(v,K_{u,v})+n_{u,s}).\label{eq: interdemiate_verf}\end{align}
By the construction of $\mathfrak{F}$, we know: \begin{align} |f(u,\varepsilon,K_{\varepsilon,u},h(m))-f(u,v,K_{u,v},h(m))|\leq 2^{r-1}.\end{align} In addition, from the rules of constructing authentication and verification polynomials, we know that $n_{\varepsilon,\mathfrak{v}}(u,K_{\varepsilon,u})\in \{0,\dots,2^{r-1}-1\}$, $n_{u,\mathfrak{a}}(v,K_{u,v})\in \{0,\dots,2^{r-2}-1\}$, and $n_{u,s}\in \{0,\dots,2^{r-2}\}$. Therefore, the verification difference $VD_{\varepsilon,u}$ must be within $\{0,\dots,2^r-1\}$.

\begin{figure}[hbt]
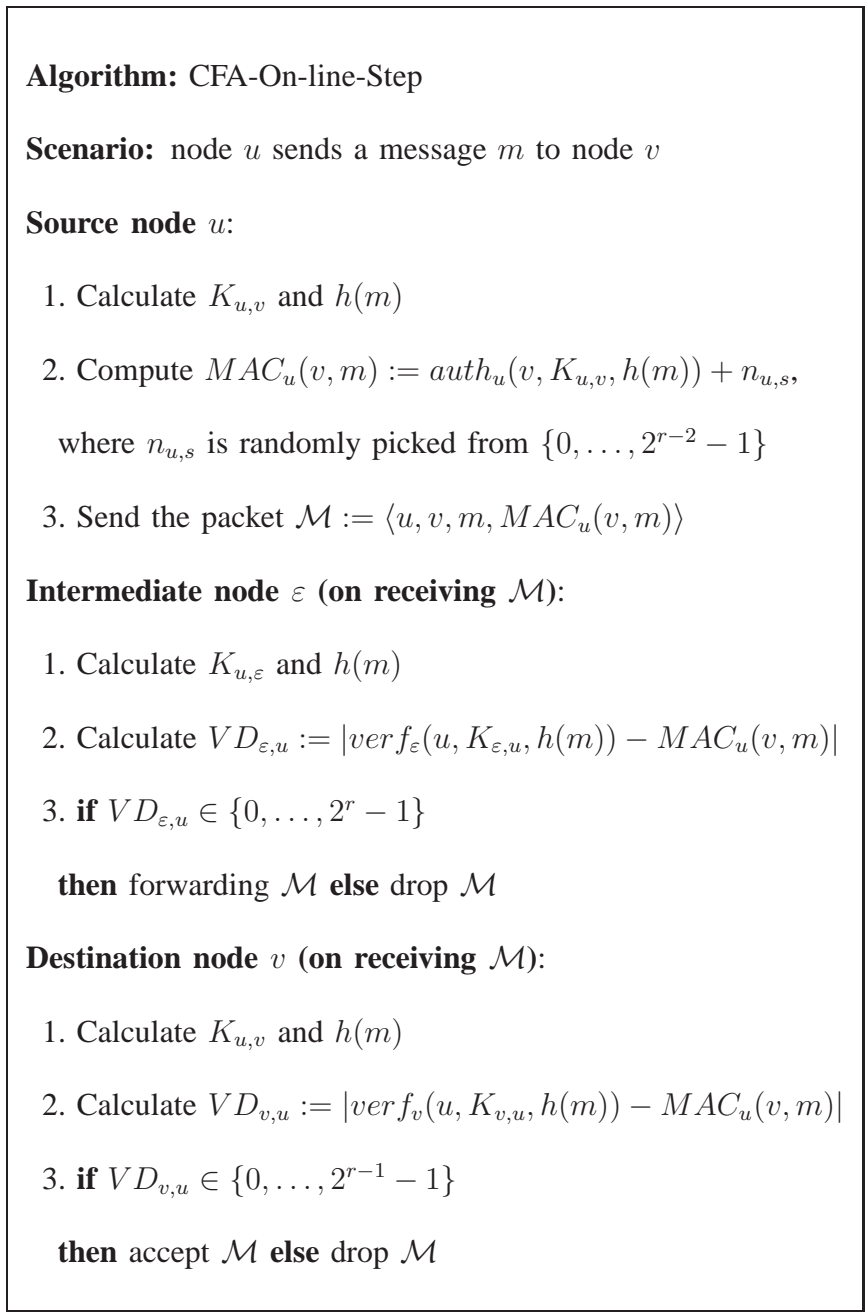

\begin{center}
\noindent  \hspace*{-0.1in} \fbox{ \parbox{10cm} {
\begin{myalgorithm}
\footnotesize
123\=123\=456\=789\=012\=345\=678\=901\=234\=567\=890\=123\=456\=789
\kill
{\bf Algorithm:} CFA-On-line-Step\\
{\bf Scenario:}\enspace node $u$ sends a message $m$ to node $v$\\
\textbf{Source node $u$}:\\
\enspace 1. Calculate $K_{u,v}$ and $h(m)$\\
\enspace 2. Compute $MAC_u(v,m):=auth_u(v,K_{u,v},h(m))+n_{u,s}$,\\
\enspace\enspace where $n_{u,s}$ is randomly picked from $\{0,\dots,2^{r-2}-1\}$\\
\enspace 3. Send the packet $\mathcal{M}:=\langle u,v,m,MAC_u(v,m)\rangle$\\
\textbf{Intermediate node $\varepsilon$ (on receiving $\mathcal{M}$)}:\\
\enspace 1. Calculate $K_{u,\varepsilon}$ and $h(m)$\\
\enspace 2. Calculate $VD_{\varepsilon,u}:=|verf_\varepsilon(u,K_{\varepsilon,u},h(m))-MAC_u(v,m)|$\\
\enspace 3. \textbf{if} $VD_{\varepsilon,u}\in \{0,\dots,2^r-1\}$ \\
\enspace\enspace \textbf{then} forwarding $\mathcal{M}$ \textbf{else} drop $\mathcal{M}$\\
\textbf{Destination node $v$ (on receiving $\mathcal{M}$)}:\\
\enspace 1. Calculate $K_{u,v}$ and $h(m)$\\
\enspace 2. Calculate $VD_{v,u}:=|verf_v(u,K_{v,u},h(m))-MAC_u(v,m)|$\\
\enspace 3. \textbf{if} $VD_{v,u}\in \{0,\dots,2^{r-1}-1\}$\\
\enspace\enspace \textbf{then} accept $\mathcal{M}$ \textbf{else} drop $\mathcal{M}$
\end{myalgorithm}
} \noindent } \caption{On-line Step of CFA.}
\label{algo: On-line Step of CFA scheme}
\end{center}
\end{figure}

On the other hand, consider the case where node $u$ has been compromised by the adversary. The adversary now wants to deceive $v$ that a message $m$ sent by $u$ is sent by $u'\neq u$. Consider the modified packet, \begin{align}\mathcal{M}'=\langle u',v,m,MAC_u(v,m)\rangle,\end{align} where $u'$ means a node ID the adversary pretends to be. Note that we only consider the adversary who exploits the information obtained from a single captured node $u$, and focus on the use of the constructed set $\mathfrak{F}$.
%The general results for security analysis will be presented in Sec. \ref{sec: Performance Evaluation}.
The verification procedure at the intermediate node $\varepsilon$ is as follows: \begin{align} &verf_\varepsilon(u',K_{\varepsilon,u'},h(m))-MAC_u(v,m)\notag\\
 =&(f(u',\varepsilon,K_{\varepsilon,u'},h(m))+n_{\varepsilon,\mathfrak{v}}(u',K_{\varepsilon,u'}))-(f(u,v,K_{u,v},h(m))+n_{u,\mathfrak{a}}(v,K_{u,v})+n_{u,s})\notag\\
    =&(f(u',\varepsilon,K_{\varepsilon,u'},h(m))-f(u,v,K_{u,v},h(m)))+(n_{\varepsilon,\mathfrak{v}}(u',K_{\varepsilon,u'})-(n_{u,\mathfrak{a}}(v,K_{u,v})+n_{u,s}).\end{align}
By the construction of $\mathfrak{F}$, we know: \begin{align} |f(u',\varepsilon,K_{\varepsilon,u'},h(m))-f(u,v,K_{u,v},h(m))|\geq 3\cdot 2^{r-1}-1.\end{align} In addition, from the construction of authentication and verification polynomials, we know that $n_{\varepsilon,\mathfrak{v}}(u',K_{\varepsilon,u'})\in \{0,\dots,2^{r-1}-1\}$, $n_{u,\mathfrak{a}}(v,K_{u',v})\in \{0,\dots,2^{r-2}-1\}$, and $n_{u,s}\in \{0,\dots,2^{r-2}\}$. Therefore, the verification difference $VD_{\varepsilon,u}$ must be not within $\{0,\dots,2^r-1\}$ and the packet $\mathcal{M}'$ will be dropped. In other words, once the source node ID of a message is modified, such malicious manipulation will be deterministically detected by the intermediate nodes. The on-line procedure of CFA is described in Fig. \ref{algo: On-line Step of CFA scheme}.

\subsection{Implementation Issues}\label{sec: Implementation Issues}
The effectiveness and efficiency of the proposed CFA scheme rely on the use of $auth_u(y,z,w)$ and $verf_u(x,z,w)$, which satisfy the constrained function set $\mathfrak{F}$, the authentication perturbation set $\mathfrak{N_a}$, and the verification perturbation set $\mathfrak{N_v}$. As the construction of $\mathfrak{N_a}$ and $\mathfrak{N_v}$ is relatively easy, in this section, we focus on the construction of $auth_u(y,z,w)$ and $verf_u(x,z,w)$, with particular emphasis on the construction of $f(x,y,z,w)$.

A straightforward method for deriving proper $f(x,y,z,w)$ is to construct the whole set $\mathfrak{F}$ and then randomly pick one from $\mathfrak{F}$. When the coefficients of the polynomials in $\mathfrak{F}$ are constrained with $\mathbb{F}_q$, there are $q^{(d+1)^4}$ possible four-variate $d$-degree polynomials. Thus, $O(q^{2\cdot(d+1)^4})$ tests are required because there are $q^{(d+1)^4}$ four-variate $d$-degree polynomials, each of which needs to check whether it satisfies the constraints $|f(x,y,z,w)-f(x',y',z',w')|\geq 3\cdot 2^{r-1}-1$, $|f(x,y,z,w)-f(x',y',z',w)|\geq 3\cdot 2^{r-1}-1$ and $|f(x,y,z,w)-f(x,y',z',w)|\leq 2^{r-1}$ in $\mathfrak{F}$, by examining the other $q^{(d+1)^4}-1$ possibilities of different input variables. The above construction of $\mathfrak{F}$ will be accomplished before sensor deployment by the network planner that is usually assumed to be resource-abundant, thus, feasible. Despite its feasibility, such an exhaustive search is not a sufficiently efficient method. In the following, we develop an efficient algorithm trading the deterministic security for the construction efficiency on the basis of the observation that, in some cases, a variant of $\mathfrak{F}$ is sufficient for our use and the search for a variant of $\mathfrak{F}$ can accelerate the construction of $f(x,y,z,w)$. Hence, we emphasize on how to efficiently construct a variant $\mathfrak{F}'$ of the original constrained function set $\mathfrak{F}$.

Let $\mathfrak{F}'$ be the \emph{weak constrained function set} as follows:
\begin{align} \mathfrak{F}'=\big\{&f(x,y,z,w)| |f(x,y,z,w)-f(x,y',z',w)|\leq 2^{r-1}, \notag \\
    &x,y\in\mathcal{I}, x'\neq x, y'\neq y, z'\neq z, w'\neq w, r<\ell\big\}.\label{eq: Fpdefinition}\end{align}

\noindent Obviously, $\mathfrak{F}$ is a subset of $\mathfrak{F}'$ since some constraints in $\mathfrak{F}$ are discarded. As to the construction of $f(x,y,z,w)$ in $\mathfrak{F}'$, our idea is to construct a random subset of $\mathfrak{F}'$ that is as large as possible and to sample a polynomial from it. Assuming that $x\in[x_{\min},x_{\max}],y\in[y_{\min},y_{\max}],z\in[z_{\min},z_{\max}]$, and $w\in[w_{\min},w_{\max}]$, we want to construct a polynomial $f(x,y,z,w)$ satisfying the constraints in $\mathfrak{F}'$, which will be shown as follows.

Assume that $f(x,y,z,w)=\sum_{i,j,k,m=0}^d\alpha_{i,j,k,m}x^iy^jz^kw^m$, $d\in \mathbb{Z}_+$, $\alpha_{i,j,k,m}\in \mathbb{F}_q$. According to the definition of $f(x,y,z,w)$ in $\mathfrak{F}'$, $f(x,y,z,w)$ can be rewritten as:
\begin{align}
\sum_{i,m=0}^d\alpha_{i,0,0,m}x^iw^m+\sum_{i,m=0,j,k=1}^d\alpha_{i,j,k,m}x^iy^jz^kw^m.
\label{eq: f'1}
\end{align}
With the representation in Eq. (\ref{eq: f'1}), the term $f(x,y,z,w)-f(x,y',z',w)$ can be written as:
\begin{align}
&\sum_{i,m=0}^d\alpha_{i,0,0,m}x^iw^m+\sum_{i,m=0,j,k=1}^d\alpha_{i,j,k,m}x^iy^jz^kw^m\notag\\
-&\left(\sum_{i,m=0}^d\alpha_{i,0,0,m}x^iw^m+\sum_{i,m=0,j,k=1}^d\alpha_{i,j,k,m}x^i(y')^j(z')^kw^m\right)\notag\\
=&\sum_{i,m=0,j,k=1}^d\alpha_{i,j,k,m}x^iw^m\left(y^jz^k-(y')^j(z')^k \right).\label{eq: f'2}
\end{align}
According to the signs of $\alpha_{i,j,k,m}$'s, Eq. (\ref{eq: f'2}) can be further rewritten as:
\begin{align}
\sum_{i,m=0,j,k=1}^d\left(\alpha_{i,j,k,m}^+x^iw^m\left(y^jz^k-(y')^j(z')^k\right)+\alpha_{i,j,k,m}^-x^iw^m\left(y^jz^k-(y')^j(z')^k \right)\right), \label{eq: f'3}
\end{align}
where
\[
\alpha_{i,j,k,m}^+=\left\{\begin{array}{ll}
\alpha_{i,j,k,m},\hfill\mbox{ if }\alpha_{i,j,k,m}>0\\
0,\hfill\mbox{ if }\alpha_{i,j,k,m}<0
\end{array} \right. \mbox{ and }
\alpha_{i,j,k,m}^-=\left\{\begin{array}{ll}
0,\hfill\mbox{ if }\alpha_{i,j,k,m}>0 \\
\alpha_{i,j,k,m},\hfill\mbox{ if }\alpha_{i,j,k,m}<0
\end{array} \right.
.\]
By taking the constraint $|f(x,y,z,w)-f(x,y',z',w)|\leq 2^{r-1}$ in $\mathfrak{F}'$ and Eq. (\ref{eq: f'3}) into consideration, we have:
\begin{align}
-2^{r-1}\leq \sum_{i,j=0,k,m=1}^d\left(\alpha_{i,j,k,m}^+x^iw^m\left(y^jz^k-(y')^j(z')^k\right)+\alpha_{i,j,k,m}^-x^iw^m\left(y^jz^k-(y')^j(z')^k \right)\right) \leq 2^{r-1}.\label{eq: f'4}
\end{align} With $-2^{r-1}$ and $2^{r-1}$ being the lower bound and upper bound of $f(x,y,z,w)-f(x,y',z',w)$, respectively, Eq. (\ref{eq: f'4}) can be rewritten as:
\begin{align} \left\{\begin{array}{ll}
\max\left\{\sum_{i,m=0,j,k=1}^d\left(\alpha_{i,j,k,m}^+x^iw^m\left(y^jz^k-(y')^j(z')^k\right)+\alpha_{i,j,k,m}^-x^iw^m\left(y^jz^k-(y')^j(z')^k \right)\right)\right\}\leq 2^{r-1}\\
\min\left\{\sum_{i,m=0,j,k=1}^d\left(\alpha_{i,j,k,m}^+x^iw^m\left(y^jz^k-(y')^j(z')^k\right)+\alpha_{i,j,k,m}^-x^iw^m\left(y^jz^k-(y')^j(z')^k \right)\right)\right\}\geq -2^{r-1}
\end{array} \right.
\label{eq: f'5}
.\end{align}
Here, we define $[f^+]=\{(i,j,k,m)|f(x,y,z,w)=\sum_{i,j,k,m=0}^d\alpha_{i,j,k,m}x^iy^jz^kw^m,\alpha_{i,j,k,m}>0\}$ and $[f^-]=\{(i,j,k,m)|f(x,y,z,w)=\sum_{i,j,k,m=0}^d\alpha_{i,j,k,m}x^iy^jz^kw^m,\alpha_{i,j,k,m}<0\}$. We can examine if a given set of $\alpha_{i,j,k,m}$'s, $\forall i,j,k,m$, constitutes a polynomial $f(x,y,z,w)$ of $\mathfrak{F}'$ by exploiting the definitions in Eq. (\ref{eq: Fpdefinition}) and considering the extremes in Eq. (\ref{eq: f'5}) shown as follows:
\begin{align} \left\{\begin{array}{ll}
&\sum_{i,m=0,j,k=1}^d\Big(\alpha_{i,j,k,m}^+x_{\max}^iw_{\max}^m\Big(y_{\max}^jz_{\max}^k-y_{\min}^jz_{\min}^k\Big)\\
    &+\alpha_{i,j,k,m}^-x_{\max}^iw_{\max}^m\Big(y_{\min}^jz_{\min}^k-y_{\max}^jz_{\max}^k \Big)\Big)\leq 2^{r-1}\\
&\sum_{i,m=0,j,k=1}^d\Big(\alpha_{i,j,k,m}^+x_{\max}^iw_{\max}^m\Big(y_{\min}^jz_{\min}^k-y_{\max}^jz_{\max}^k\Big)\\
    &+\alpha_{i,j,k,m}^-x_{\max}^iw_{\max}^m\Big(y_{\max}^jz_{\max}^k-y_{\min}^jz_{\min}^k \Big)\Big)\geq -2^{r-1}
\end{array} \right.
\label{eq: f'6}
.\end{align}
Define $f'(x,y,z,w)=\sum_{i,j,k,m=0}^d\alpha'_{i,j,k,m}x^iy^jz^kw^m$, which is only different from $f(x,y,z,w)$ in the part of coefficients. From Eq. (\ref{eq: f'6}), we can observe that the possible range of $|f'(x,y,z,w)-f'(x,y',z',w)|$ will be contained in $|f(x,y,z,w)-f(x,y',z',w)|$, \emph{i.e.}, $\max\{f'(x,y,z,w)-f'(x,y',z',w)\}\leq \max\{f(x,y,z,w)-f(x,y',z',w)\}$ and $\min\{f'(x,y,z,w)-f'(x,y',z',w)\}\geq \min\{f(x,y,z,w)-f(x,y',z',w)\}$, if (i) $\alpha_{i,j,k,m}-\alpha'_{i,j,k,m}\geq 0$, $\forall (i,j,k,m)\in[f^+]$, or (ii) $\alpha_{i,j,k,m}-\alpha'_{i,j,k,m}\leq 0$, $\forall (i,j,k,m)\in[f^-]$. With this \emph{monotone} property, our algorithm, randomly sampling a polynomial from a random subset of $\mathfrak{F}'$, whose pseudo code is shown in Fig. \ref{algo: F' Construction}, can be described as follows.

\begin{figure}[hbt]
\begin{center}
\noindent  \hspace*{-0.1in} \fbox{ \parbox{10cm} {
\begin{myalgorithm}
\footnotesize
123\=123\=456\=789\=012\=345\=678\=901\=234\=567\=890\=123\=456\=789
\kill
{\bf Algorithm:} $\mathfrak{F}'$-Construction ($[\alpha]$ \texttt{is the final output of this algorithm})\\
1.\quad randomly select $[\alpha]$ and construct $[f^+]$ and $[f^-]$\\
2.\quad \bwhile $[\alpha]$ cannot satisfy Eq. (\ref{eq: f'6})\\
3.\quad\quad $[\alpha]:=[\alpha/2]$\\
4.\quad randomly construct $\Omega:=\{(i,j,k,m)|0\leq i,j,k,m\leq d\}$ with $|\Omega|\geq 0$, and set $[\phi]=[\alpha]$\\
5.\quad \bfor each element $(i,j,k,m)$ in $\Omega$\\
6.\quad\quad \bif $(i,j,k,m)\in[f^+]$\\
7.\quad\quad\quad find the maximum $\varphi$ such that $\langle[\alpha],(i,j,k,m),\varphi\rangle$ is satisfied with Eq. (\ref{eq: f'6})\\
8.\quad\quad\quad $\phi$ is randomly selected from $[0,\varphi]$, and set $[\alpha]:=\langle[\alpha],(i,j,k,m),\phi\rangle$\\
9.\quad\quad \bif $(i,j,k,m)\in[f^-]$\\
10.\quad\quad\quad find the minimum $\varphi$ such that $\langle[\alpha],(i,j,k,m),\varphi\rangle$ is satisfied with Eq. (\ref{eq: f'6})\\
11.\quad\quad\quad $\phi$ is randomly selected from $[0,\varphi]$, and set $[\alpha]:=\langle[\alpha],(i,j,k,m),\phi\rangle$\\
\end{myalgorithm}
} \noindent } \caption{$\mathfrak{F}'$-Construction algorithm}
\label{algo: F' Construction}
\end{center}
\end{figure}

As $\alpha_{i,j,k,m}$ in $f(x,y,z,w)$ denotes the coefficient of $x^iy^jz^kw^m$ for specified $i,j,k,m$, we use $[\alpha]$ to denote an instance of $\alpha_{i,j,k,m}$'s, $\forall i,j,k,m$. At the beginning of $\mathfrak{F}'$-Construction algorithm shown in Fig. \ref{algo: F' Construction}, we randomly choose $[\alpha]$ and determine if the chosen $[\alpha]$ satisfies Eq. (\ref{eq: f'6}). If $[\alpha]$ fails to satisfy Eq. (\ref{eq: f'6}), $[\alpha]=[\alpha/2]$ is checked recursively until Eq. (\ref{eq: f'6}) is satisfied (Lines 1$\sim$3). Here, $[\alpha/2]$ consists of $\lfloor\frac{\alpha_{i,j,k,m}}{2}\rfloor$'s, where each $\alpha_{i,j,k,m}$ is an element in $[\alpha]$. Note that the loop (Lines 2$\sim$3) is guaranteed to terminate at a certain step because at least the setting of $\alpha_{i,j,k,m}=0$, $\forall i,j,k,m$, is satisfiable. With the monotone property, we can also guarantee that any polynomial sampling from $\{[\alpha']|[\alpha']\preceq[\alpha]\}$ is one of the polynomials in $\mathfrak{F}'$. Here, $[\alpha']\preceq[\alpha]$ means that the possible range of $|f'(x,y,z,w)-f'(x,y',z',w)|$ will be contained in $|f(x,y,z,w)-f(x,y',z',w)|$. Thus, after the execution of Line 3, we can sample a polynomial $f(x,y,z,w)\in \mathfrak{F}'$ from the sample space $\{[\alpha']|[\alpha']\preceq[\alpha]\}$. Nevertheless, we can, in fact, further extend the sample space by tuning selected $\alpha_{i,j,k,m}$'s (Lines 5$\sim$11). For example, suppose $|\Omega|$ $\alpha_{i,j,k,m}$'s are chosen to be tuned. In particular, defining $\langle[\alpha],(i,j,k,m),\varphi\rangle$ as $[\alpha]$ whose $\alpha_{i,j,k,m}$ is selected to be replaced by $\varphi$, we can extend the range of $|f'(x,y,z,w)-f'(x,y',z',w)|$ by maximizing (minimizing) the selected $\alpha_{i,j,k,m}$ if $(i,j,k,m)\in[f^+]$ ($(i,j,k,m)\in[f^-]$) so that the size of $\{[\alpha']|[\alpha']\preceq[\alpha]\}$ will be increased. Together with $[\alpha]$ obtained after Line 3, Lines 8 and 11 behave like sampling a polynomial from a subset of $\mathfrak{F}'$, which could be randomly different due to the random construction of $\Omega$ (Line 4). Note that a search of maximum $\overline{\alpha}_{i,j,k,m}$ (Line 8) can be accomplished by conducting binary search on the positive integers greater than $\alpha_{i,j,k,m}$. The minimum $\overline{\alpha}_{i,j,k,m}$ can be found in a similar way (Line 11). Since we should conduct binary search once for each element in $\Omega$, the running time of $\mathfrak{F}'$-Construction algorithm is $O(|\Omega|\log q)$. Indeed, from the theoretical point of view, it might obtain only a useless constant polynomial after the execution of $\mathfrak{F}'$-Construction algorithm, therefore, require executing the algorithm multiple times. Nevertheless, in practice, when a sufficiently large security parameter $r$ (as defined in Eq. (\ref{eq: Fdefinition}) and Eq. (\ref{eq: Fpdefinition})) is selected, executing the algorithm once is sufficient for sampling a non-trivial polynomial from $\mathfrak{F}'$. It should be noted that $\mathfrak{F}'$-Construction algorithm is not a uniform sampling over $\mathfrak{F}'$. As we mentioned earlier, what we do is to construct and then sample from a random subset of $\mathfrak{F}'$. Nevertheless, due to the use of $\Omega$ with the purpose of tuning randomly selected $\alpha_{i,j,k,m}$'s, we can still guarantee that there is a nonzero probability of each polynomial in $\mathfrak{F}'$ being sampled, resulting the sufficient security against directly guessing all the coefficients $\alpha_{i,j,k,m}$'s.

When $f(x,y,z,w)$ is selected from the weak constrained function set, $\mathfrak{F}'$, the filtering capability will be slightly reduced. Its impact on the security of CFA using $f(x,y,z,w)\in \mathfrak{F}'$ is discussed in the following. Even if $f(x,y,z,w)$ is selected from $\mathfrak{F}'$, the destination and the intermediate nodes, when receiving the genuine message, can still correctly accept and forward the received message, respectively. The validation procedures are the same as those in Eqs. (\ref{eq: destination_verf}) and (\ref{eq: interdemiate_verf}), therefore, are omitted here. The destination node and intermediate nodes, however, only \emph{probabilistically} drop falsified messages in CFA using $f(x,y,z,w)\in \mathfrak{F}'$, instead of deterministically dropping the modified messages in CFA using $f(x,y,z,w)\in \mathfrak{F}$. The principle behind this change is as follows:
\begin{align} &verf_\varepsilon(u',K_{\varepsilon,u'},h(m))-MAC_u(v,m)\notag\\
 =&(f(u',\varepsilon,K_{\varepsilon,u'},h(m))+n_{\varepsilon,\mathfrak{v}}(u',K_{\varepsilon,u'}))-(f(u,v,K_{u,v},h(m))+n_{u,\mathfrak{a}}(v,K_{u,v})+n_{u,s})\notag\\
    =&(f(u',\varepsilon,K_{\varepsilon,u'},h(m))-f(u,v,K_{u,v},h(m)))+(n_{\varepsilon,\mathfrak{v}}(u',K_{\varepsilon,u'})-(n_{u,\mathfrak{a}}(v,K_{u,v})+n_{u,s})).\label{eq: Fpsecurity}
\end{align}
The first term $(f(u',\varepsilon,K_{\varepsilon,u'},h(m))-f(u,v,K_{u,v},h(m)))$ on the RHS of Eq. (\ref{eq: Fpsecurity}) can be an arbitrary element in $\mathbb{F}_q$, leading also to the arbitrariness of the final result in Eq. (\ref{eq: Fpsecurity}). Therefore, the probability that $|verf_\varepsilon(u',K_{\varepsilon,u'},h(m))-MAC_u(v,m)|$ happens to be within the range $[-2^r+1,2^r-1]$ is increased from $0$ to $\frac{2^{r+1}-1}{q}$. With a similar argument, one can also show that the probability of detecting falsified messages is $\frac{2^r-1}{q}$.

\section{CFA-Based En-Route Filtering Scheme}\label{sec: CFA-Based En-Route Filtering Scheme}
With CFA described in Sec. \ref{sec: The Constrained Function Based Authentication (CFA) Scheme}, the design of CFA-based en-route filtering (CFAEF) scheme is straightforward. The CFAEF scheme consists of three phases: node initialization phase, report endorsement phase, and en-route filtering phase, which, respectively, will be described as follows.

\textbf{Node initialization phase.} At first, a global security parameter $t$, which indicates the maximum number of compromised nodes tolerable in the CFAEF scheme, is selected. If the number of compromised nodes exceeds $t$, then the adversary can inject falsified data without being detected. It should be noted that such a limitation is also applied to all en-route filtering schemes unless additional location information is used. In addition, each node $u$ is preloaded with $auth_u(y,z,w)$ and $verf_u(x,z,w)$ prepared for the use of CFA. Last, the sensor nodes are deployed on the sensing region.

\textbf{Report endorsement phase.} After sensor deployment, a node enters this phase when it has an event report to be sent\footnote{An event could be simultaneously observed by multiple nodes. Here we assume that one of these detecting nodes is responsible for sending the event report, but the election of such node is beyond the scope of this paper.}. More specifically, once a node $u$ wants to send an event report $E$ to a destination node $v$, it first broadcasts $E$ in plaintext to the nodes neighboring to $u$. If the neighboring node $\mu$ agrees with $E$, then it generates a MAC, $MAC_\mu(v,E)$ via the proposed CFA scheme, and sends an endorsement of $E$, $MAC_\mu(v,E)$, back to $u$. After collecting $t$ MACs from the neighboring nodes\footnote{The WSNs in our consideration possess high node density such that $t$-coverage \cite{ht05,wcp06,wsc08} can be achieved.}, $\mu_1,\dots,\mu_t$, $u$ first checks whether the value of $|verf_u(\mu_j,E)-MAC_{\mu_j}(v,E)|$, $j=1,\dots,t$, is within the predetermined range $[0, 2^r-1]$. If some of the collected endorsements, $MAC_{\mu_j}(v,E)$, fail to be verified, $u$ drops all $MAC_{\mu_j}(v,E)$'s and acquires other endorsements from the neighboring nodes other than $\mu_1\dots,\mu_t$. Only when all of $t$ collected endorsements are successfully verified, $u$ forwards $\langle E,u,v,MAC_u(v,E),\mu_1,MAC_{\mu_1}(v,E),\dots,\mu_t,MAC_{\mu_t}(v,E)\rangle$ to $v$.

\textbf{En-route filtering phase.} Once receiving the packet \[(E,u,v,MAC_u(v,E),\mu_1,MAC_{\mu_1}(v,E),\dots,\mu_t,MAC_{\mu_t}(v,E)),\] the intermediate node $\varepsilon$ first checks whether the attached endorsements are generated by $t+1$ distinct nodes. The packet is dropped if the verification fails. Afterwards, for each $\nu$ of the $t+1$ endorsements, node $\varepsilon$ checks whether $VD_{\varepsilon,\nu}=|verf_\varepsilon(\nu,E)-MAC_{\nu}(v,E)|$ is within the predetermined range $[0, 2^r-1]$. Only if node $\varepsilon$ succeeds in verifying all the $t+1$ endorsements, is the packet forwarded. Otherwise, the packet is dropped. The operation performed by the destination node $v$ is similar to that performed by the intermediate node. The difference is that $v$ checks whether $VD_{v,\nu}=|verf_v(\nu,E)-MAC_{\nu}(v,E)|$ is within the predetermined range $[0, 2^{r-1}-1]$. Only if $v$ succeeds in verifying all the $t+1$ endorsements, is the event report $E$ accepted. Otherwise, the packet is dropped.

\section{Performance and Security Evaluation}\label{sec: Performance and Security Evaluation}
In this section, for CFAEF, in addition to analyzing the overhead (Sec. \ref{sec: Overhead Analysis}), we study its security (Sec. \ref{sec: Security}) and compare the energy saving with the other methods (Sec. \ref{sec: Energy Savings}).

\subsection{Overhead Analysis}\label{sec: Overhead Analysis}
As to the storage overhead, two trivariate polynomials need to be stored in each node in CFA, as shown in Fig. \ref{algo: Off-line Step of CFA scheme}. Therefore, in CFAEF, the storage overhead $O(d^3)$ is required due to the use of authentication and verification polynomials.

For the endorsing node, the computation overhead comes from the calculation of the message authentication code, which involves trivariate polynomial evaluation and requires $O(d^3)$ arithmetic operations \cite{ck04,sc05}. On the other hand, the computation overhead for the source node, intermediate nodes, and destination node is the same, which is $O(td^3)$, because $t$ MACs should be calculated.

As to the communication overhead of CFAEF, the source node has to communicate with the neighboring nodes to obtain the endorsements. Moreover, the source node has to send $\langle E,u,v,MAC_u(v,E),\mu_1,MAC_{\mu_1}(v,E),\dots,\mu_t,MAC_{\mu_t}(v,E)\rangle$, instead of $\langle E,u,v\rangle$, to the destination node. As a result, the additional communication overhead incurred by the use of CFAEF is $O(tH)$, where $H$ is the average number of hops between two arbitrary nodes in a network.

\subsection{Security}\label{sec: Security}
First, we study the security of the proposed CFA scheme. In particular, we assume that the adversary attempts to recover the coefficients of $f(x,y,z,w)$. Consider an adversary who can only modify the transmitted packet and re-transmit the modified one in order to deceive the destination node into accepting that the packet originates from the other node or that the message is authentic. The probability of the adversary successfully deceiving the destination node can be analyzed as follows.
If the message $m$ with $MAC_u(v,m)$ sent by the node $u$ is modified to $m'\neq m$ or $u'\neq u$, then we can know that the probability that the intermediate node forwards the message $m'$ is at most $\frac{2^{r+1}-1}{q}$ and the probability that the destination node accepts the message $m'$ is at most $\frac{2^{r}-1}{q}$. This can be explained by the fact that, to deceive the destination node, the best strategy that can be adopted by the adversary is to forge the MAC corresponding to $m'$ and $u'$. Nonetheless, such MAC can only be guessed by the adversary. Therefore, the verification difference would be arbitrary and the probabilities that $VD_{\varepsilon,u}$ and $VD_{v,u}$ happen to be within the predetermined ranges are at most $\frac{2^{r+1}-1}{q}$ and $\frac{2^{r}-1}{q}$ for the intermediate node and destination node, respectively.

Second, we consider the case where the adversary not only eavesdrops on the transmitted messages but also compromises $n$ nodes to use the security information stored in them, trying to recover the coefficients of $f(x,y,z,w)$. We can know that the adversary cannot break $f(x,y,z,w)$ if only $n\leq d$ nodes are compromised \cite{bshkvy93}. When the adversary has compromised $n>d$ nodes, the complexity for it to obtain the coefficients of $f(x,y,z,w)$ is $\Omega(q^{d+1})$. This can be explained as follows. Assume that $u_0,\dots,u_{n-1}$ are $n$ compromised nodes. Let $x_0$, $z_0$, and $w_0$ be arbitrary elements in $\mathbb{F}_q$. We know that if we can arbitrarily construct $f(x_0,y,z_0,w_0)$ for any $x_0$, $y_0$, and $w_0$, then the coefficients of $f(x,y,z,w)$ can be inferred by solving a system of equations. Thus, our goal is to obtain the coefficients of $f(x_0,y,z_0,w_0)$. Note that the discussion and effect of obtaining the coefficients of, for example, $f(x,y_0,z_0,w_0)$ is the same as that of obtaining the coefficients of $f(x_0,y,z_0,w_0)$. Thus, we omit the former case and focus only on the latter case here. We can know that $f(x_0,y,z_0,w_0)$ can always be written as $\sum_{j=0}^{d}C_jy^j$. Based on the construction of $verf_u(x,z,w)$, we can derive the following $n$ equations:
\begin{align}
\sum_{j=0}^{d}C_j(u_i)^j=verf_{u_i}(x_0,z_0,w_0)-n_{u_i,\mathfrak{v}}(u_i,z_0), 0\leq i\leq n-1.
\label{eq: security proof 1}
\end{align}
In this system of equations, there are $d+1+n$ unknown variables including $C_j$ ($0\leq j\leq d$) and $n_{u_i,\mathfrak{v}}(u_i,z_0)$ ($0\leq i\leq n-1$). There are, however, only $n$ equations. Thus, $d+1$ unknown variables should be eliminated or correctly guessed. The polynomials, $auth_{u_i}(y,z,w)$'s, may be used by the adversary to reduce the number of unknown variables. A common method that is able to reduce the number of unknown variables is called \emph{reflection attack} in \cite{zsw08} and is employed here. Let $a_i=verf_{u_i}(u_0,z_0,w_0)-auth_{u_0}(u_i,z_0,w_0)=n_{u_i,\mathfrak{v}}(u_i,z_0)-n_{u_0,\mathfrak{a}}(u_i,z_0)$. The above equation can be rewritten as $n_{u_i,\mathfrak{v}}(u_i,z_0)=a_i+n_{u_0,\mathfrak{a}}(u_i,z_0)$. Together with this equation, Eq. (\ref{eq: security proof 1}) can be represented as:
\begin{align}
\sum_{j=0}^{d}C_j(u_i)^j=verf_{u_i}(x_0,z_0,w_0)-a_i-n_{u_0,\mathfrak{a}}(u_i,z_0), 0\leq i\leq n-1.
\label{eq: security proof 2}
\end{align}
It can be observed that reflection attack does not work in breaking $f(x,y,z,w)$ with higher probability because there are still $d+1+n$ unknown variables in $n$ equations. Thus, $d+1$ unknown variables should be eliminated or correctly guessed. Since each unknown variable can be of at least $r$ bits length, the complexity of recovering the coefficients is $\Omega(2^{r(d+1)})$.

After the security of CFA is established, the resilience of CFAEF against false data injection attack, PDoS attack, and FEDoS attack is obvious. For example, CFAEF is resilient to false data injection attack and PDoS attack because, with the MACs generated by CFA, the false data can be detected and dropped by either intermediate nodes or the destination node when the number of compromised nodes does not exceed $t$. In particular, with $t$ endorsements required, the probabilities of detecting the bogus message on each intermediate node and the destination node are $(\frac{2^{r+1}-1}{q})^t$ and $(\frac{2^r-1}{q})^t$, respectively. On the other hand, FEDoS attack is useless because if the compromised node sends a false endorsement to the source node, the source node, acting as the intermediate node between the endorsing node and destination, can identify the false endorsement via the CFA verification, and refuse to communicate with the compromised node thereafter.

\subsection{Energy Savings}\label{sec: Energy Savings}
In this section, the energy consumption model similar to that used in \cite{yg06} is used to analyze the energy savings of various schemes. Due to the fact that, the higher the filtering capability, the lower the energy consumed for forwarding falsified messages, the evaluation of energy consumption is somewhat equivalent to the evaluation of the filtering capability.

As described in Sec. \ref{sec: Introduction}, most of the existing en-route filtering schemes require strict assumptions. For example, IHA \cite{zsjn04} and GREF \cite{yl09} heavily rely on the sophisticated security association that must be established within a period of secure bootstrapping time, which is unrealistic in certain cases. Moreover, some schemes \cite{rlz06,yl09,yyyla05} require location information and some others \cite{ksbe07,yl04,yzzm08} work only on query-based networks. Therefore, in the following, we emphasize the energy consumption comparison among SEF \cite{yllz04}, DEF \cite{yg06}, and our CFAEF scheme, because SEF and DEF achieve the balance among efficiency, filtering capability, and generality while minimal assumptions are required.

In particular, \cite{yg06} shows a general formula
\begin{align}
E=L_r(H+\frac{\beta}{p}), \label{eq: energy consumption formula}
\end{align}
for evaluating the energy consumption $E$ of report forwarding. Here, $L_r$, $H$, $\beta$, and $p$ denote the bit-length of the report plus endorsements, the average number of hops between two arbitrary nodes, the ratio of the false report to the legitimate report, and the probability of detecting the false report on each node, respectively. Note that, as demonstrated in \cite{ba06}, the energy consumed by the communication is over $1000$ times greater than that consumed by the computation. Thus, the above formula emphasizes on the calculation of energy consumption incurred by the communications. Let $e_{Ord}$ be the energy consumption of report forwarding when no filtering scheme is used. Let $e_{SEF}$ and $e_{DEF}$ be the energy consumption of report forwarding when SEF \cite{yllz04} and DEF \cite{yg06} are used, respectively. Throughout the energy evaluation, the common parameters $t=5$, $\beta=10$, MAC size, $64$ bits, and the byte-length of the report, $24$ bytes, were used for the methods adopted for comparisons. For \cite{yg06}, with default parameter settings, we know that $e_{Ord}=2112H$, $e_{SEF}=306(H+200)$, and $e_{DEF}\cong 732(H+36)$. According to Eq. (\ref{eq: energy consumption formula}), with similar calculation\footnote{In \cite{yg06}, the packet length is only calculated based on counting the lengths of the report and MACs excluding the lengths contributed from the source node ID, destination node ID, and endorsing nodes IDs.} to \cite{yg06} and the setting of $q=127$ and $r=120$, the energy consumption $e_{CFAEF}$ in CFAEF can also be derived as $e_{CFAEF}\cong 512(H+10)$, because in this case $p\cong 1$ and $L_r=24\times 8+5\times 64=512$. In particular, when $H=50$, our scheme saves $1-\frac{E_{CFAEF}}{E_{Ord}}\cong 71\%$ of energy than the  scheme without using filtering, $1-\frac{E_{CFAEF}}{E_{SEF}}\cong 60\%$ of energy than SEF, and $1-\frac{E_{CFAEF}}{E_{DEF}}\cong 51\%$ of energy than DEF.

\section{Conclusion}
A Constrained Function based message Authentication (CFA) scheme, which can be thought of as a hash function directly supporting en-route filtering functionality, is proposed. According to CFA, we construct a CFA-based En-route Filtering (CFAEF) scheme to simultaneously defend against false data injection, PDoS, and FEDoS attacks. Some theoretical and numerical analyses are provided to demonstrate the efficiency and effectiveness of CFAEF.\\

\noindent
{\bf Acknowledgment\/}:
Chia-Mu Yu and Chun-Shien Lu were supported by NSC 97-2221-E-001-008. Sy-Yen Kuo was supported by NSC 96-2628-E-002-138-MY3.

\end{document}